\documentclass[12pt,preprint]{aastex}

\slugcomment{To appear in The Astrophysical Journal}

\shorttitle{FIR Features in ISM Emission}
\shortauthors{Onaka \& Okada}

\begin{document}


\title{Detection of Far-Infrared Features in Star-Forming 
Regions\altaffilmark{1}}


\author{Takashi Onaka and Yoko Okada}
\affil{Department of Astronomy, Graduate School of Science, University of 
Tokyo,
Bunkyo-ku, Tokyo 113-0033, Japan}
\email{onaka@astron.s.u-tokyo.ac.jp; okada@astron.s.u-tokyo.ac.jp}


\altaffiltext{1}{Based on observations with ISO, an ESA project with 
instruments funded by ESA members states (especially the PI countries France, 
Germany, the Netherlands, and the United Kingdom) and with the participation
of ISAS and NASA.}


\begin{abstract}
We report the detection of a feature at 65\,$\mu$m and a broad feature 
around 100\,$\mu$m
in the far-infrared spectra of the diffuse emission from two active star-forming 
regions, \objectname{the Carina nebula} and \objectname
{Sharpless 171}.  The features are seen in the spectra over a wide area of the 
observed regions, indicating that the carriers are fairly ubiquitous
species in the interstellar medium. 
A similar 65\,$\mu$m feature has been detected in evolved stars and  
attributed to diopside, a Ca-bearing crystalline silicate.
The present observations indicate the first detection of a crystalline 
silicate in the interstellar medium
if this identification holds true also for the interstellar feature.  
A similar broad feature around 90\,$\mu$m reported in the spectra
of evolved stars has been attributed to calcite, a Ca-bearing carbonate mineral.
The interstellar feature seems to be shifted to longer wavelengths
and have a broader width although the precise estimate of the feature
profile is difficult.
As a carrier for the interstellar 100\,$\mu$m 
feature, we investigate the possibility that
the feature originates from carbon onions, grains consisting of
curved graphitic shells.  Because of the curved graphitic sheet structure, the 
optical properties
in the direction parallel to the graphitic plane interacts with those in the 
vertical 
direction in carbon onion grains.  This effect enhances the
interband transition feature in the direction parallel to the graphitic plane in 
carbon onions, 
which is suppressed in graphite particles.  Simple calculations suggest
that carbon onion grains are a likely candidate for the observed 100\,$\mu$m 
feature carrier, but the appearance of the
feature is sensitive to the assumed optical properties.
\end{abstract}




\section{Introduction}

Infrared spectroscopy provides a crucial means in the identification of
interstellar dust compositions.  Recent infrared satellite observations by
the {\it Infrared Telescope in Space} \citep[{\it IRTS};][]{mur96} and the 
{\it Infrared Space 
Observatory} \citep[{\it ISO};][]{kes96} have
revealed several new dust features in the diffuse emission, indicating the
presence of new dust components in the interstellar medium \citep{ona96, mat96,
chan2000}.  Observations
by {\it ISO} also clearly show the presence of 
crystalline
silicates around young and evolved stars for the first time \citep{wat96, wae96, 
kemp2002, mol2002}, while it is not yet certain whether crystalline silicates 
exist commonly in interstellar space.  

In the present paper we report the observations of active star-forming regions,
the Carina nebula and the Sharpless 171 (S171) region 
with the Long-Wavelength Spectrometer
\citep[LWS;][]{clegg96} on board {\it ISO} and the detection of far-infrared  
features around 65\,$\mu$m and 100\,$\mu$m in
the diffuse emission.  The Carina nebula is one of the most active regions
on the Galactic plane and known to contain a number of early-type stars
\citep{wal95, fei95}.  The S171 region is a typical \ion{H}{2} region and
molecular cloud complex \citep{yan92}.  Both regions are supposed to 
represent the characteristics of active regions in the Galaxy.
Possible carriers of the two detected features are discussed and we 
investigate the possibility that carbon onion
grains of curved graphitic shells are the carrier for the broad interstellar
100\,$\mu$m feature.

\section{Observations and Results}
The central $40\arcmin \times 20\arcmin$ portion of the Carina nebula 
was observed by two-dimensional raster scans 
with the LWS full grating
scan mode and the far-infrared spectra for 43--197\,$\mu$m were obtained
for 132 positions
\citep[for details of the observations, see][]{miz2002}.  A one-dimensional
scan was made for 24 positions on a line
from the heating source to the molecular cloud region
in S171 with the same LWS observing mode \citep{oka2002}.  The observed
area of the both objects includes ionized regions and molecular clouds
and the spectra sample the diffuse emission from the interstellar matter
rather than the emission from point-like sources.
The Off-Line Processing
data of version 10.1 (OLP 10.1) provided by the ISO Archival Data Center
were used in the present study.  The
spectra were defringed, converted into the surface brightness, and
the extended source correction was applied
by the ISO Spectral Analysis Package (ISAP\footnote
{The ISO Spectral Analysis Package (ISAP) is a joint development by the LWS 
and SWS Instrument Teams and Data Centers. Contributing institutes are
CESR, IAS, IPAC, MPE, RAL and SRON.}) software.  The beam size and the
correction factors were taken from the latest LWS Handbook
\citep{gry2002}.  There are gaps in the spectra between the detector channels,
which can be ascribed to the uncertainties either in the responsivity,
in the dark current, or in the spatial brightness distribution in the beam.

\begin{figure}
\epsscale{0.85}
\plotone{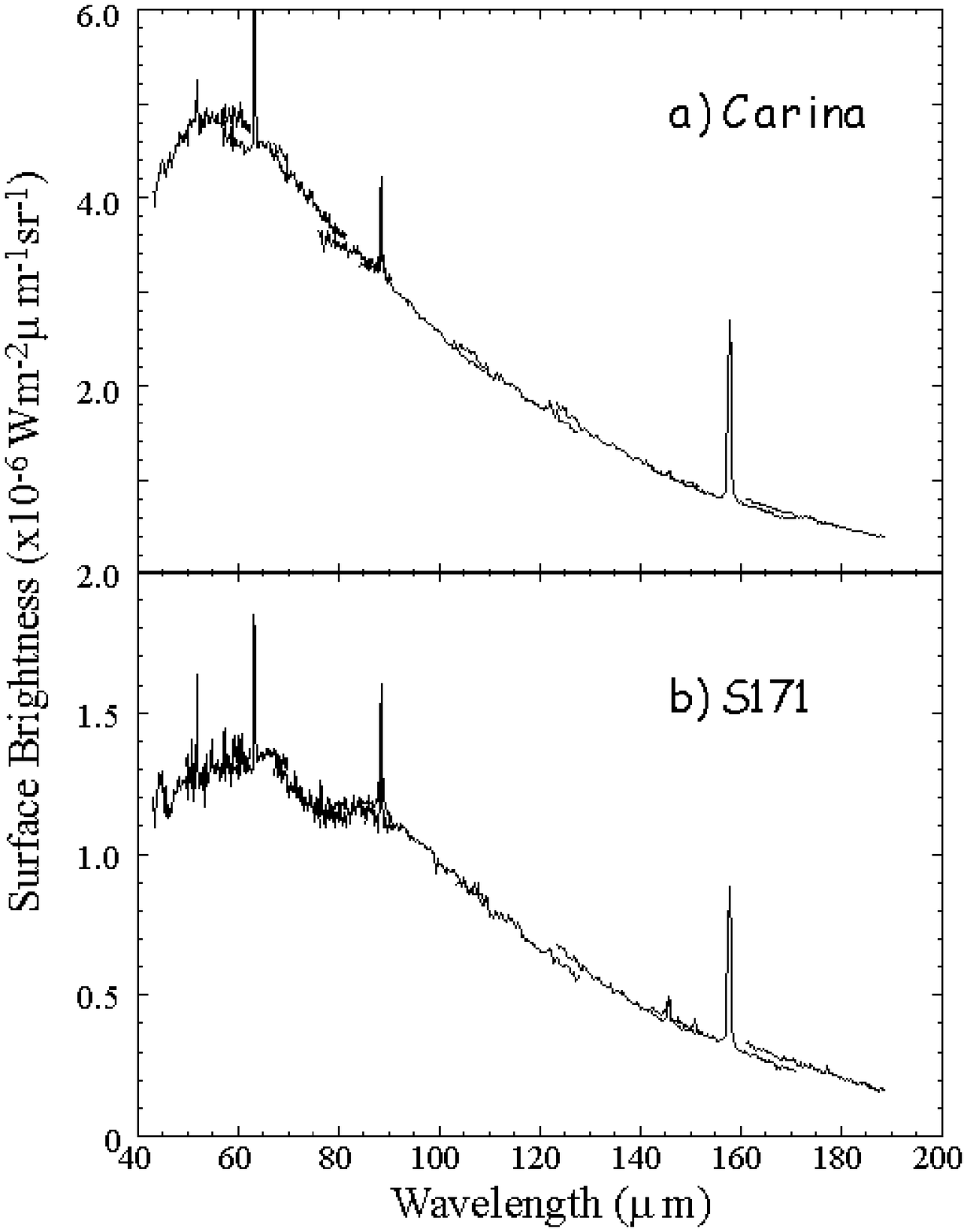}
\caption{Examples of unstitched LWS
spectra of the Carina nebula (a) and
the Sharpless 171 (S171) (b).  The Carina nebula spectrum was taken at
$(l, b) = (287.31\degr, -0.69\degr)$, while the S171 spectrum at
$(l, b) = (117.95\degr, 4.98\degr)$. 
Strong sharp lines are [\ion{O}{1}]63\,$\mu$m, 
[\ion{O}{3}]88\,$\mu$m, and 
[\ion{C}{2}]158\,$\mu$m. \label{fig1}}
\end{figure}

\begin{figure}
\epsscale{0.79}
\plotone{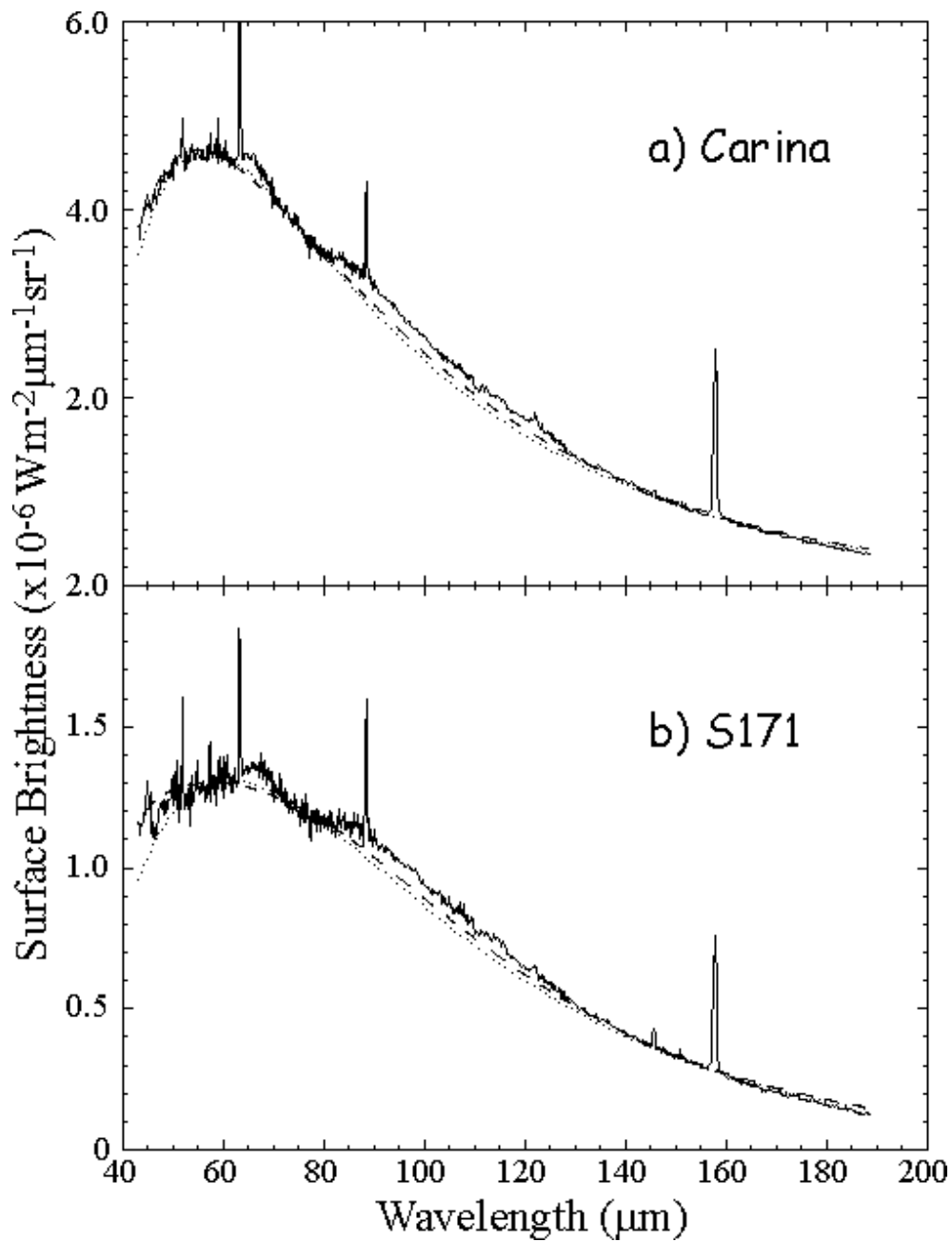}
\caption{Stitched spectra of the Carina nebula (a) and the Sharpless
171 (b) of Figure~\ref{fig1}.  The best fit continua of the
astronomical silicate and graphite model are also indicated.
The dotted lines indicate the fitted continuum to  140--190\,$\mu$m (case A), 
while the dashed lines indicate that to 120--190\,$\mu$m (case B; see text).
The temperatures of silicate and graphite grains are, 21.7K and 39.2K,
23.3K and 41.1K for the Carina and S171 regions for case A, respectively.
They are 23.0K, 42.0K for the Carina region, and 24.4K and 49.9K for the
S171 region for case B.
\label{fig2}}
\end{figure}

Figure~\ref{fig1} shows examples of the obtained original
spectra of the two regions, while Figure~\ref{fig2} indicates their stitched
spectra to correct the gaps.  Both
spectra were taken at the interface regions between the molecular cloud and 
ionized gas,
where the far-infrared intensity is sufficiently large to investigate 
dust features.  The stitched spectra are made
by scaling each detector signal because the observed
regions are bright enough that the uncertainty in the dark current should be
less significant than those in the responsivity or in the spatial brightness
distribution.  As can be seen in Figure~\ref{fig1}, the amount of the gaps is 
small ($< \pm$ 5\%) except for the three longest channels ($> 120$\,$\mu$m),
where 10--20\% scaling is necessary to correct the gaps.
The presence of a relatively narrow band feature at 
65\,$\mu$m is seen even in the unstitched spectra, particularly in S171.
In the spectrum of the Carina nebula, the appearance of this
feature is slightly disturbed by
the higher levels of the adjacent channel spectra (SW2 and SW4) relative to 
the level of the SW3 channel, but it can still be
seen in the individual spectrum of the SW3 channel. 
A broad feature centered around 100\,$\mu$m is also noticeable in the unstitched 
spectrum of S171.  The slope of the
continuum starts to become flatter around the boundary between the SW4 and SW5 
channels, indicating a feature starting around 80\,$\mu$m.
In the unstitched spectrum of the Carina nebula, the gap between the SW4 and
SW5 channels makes the feature less obvious, but the change in the slope in 
the SW5 channel 
can still be seen.  The stitched spectrum 
clearly indicates the presence of the feature.  
However there is no appreciable abrupt change 
in the slope at longer wavelengths and the longer wavelength end of the feature 
is difficult to estimate from these spectra.  
Neither spurious features have been reported nor the relative spectral
response functions have the corresponding features in these spectral ranges
\citep{gry2002}.
We will discuss possible 
underlying
continua to confirm the presence of the feature and estimate the
feature profile in next section.  
Similar
features are seen at about a half of the observed positions both in the
Carina and S171 regions.  Since these features are seen in a 
wide area of the interstellar medium, the band carriers must be 
ubiquitous species in interstellar space.

\placefigure{fig1}

\placefigure{fig2}

\section{Discussion}
\subsection{Interstellar 65\,$\mu$m feature}
\begin{figure}
\epsscale{0.85}
\plotone{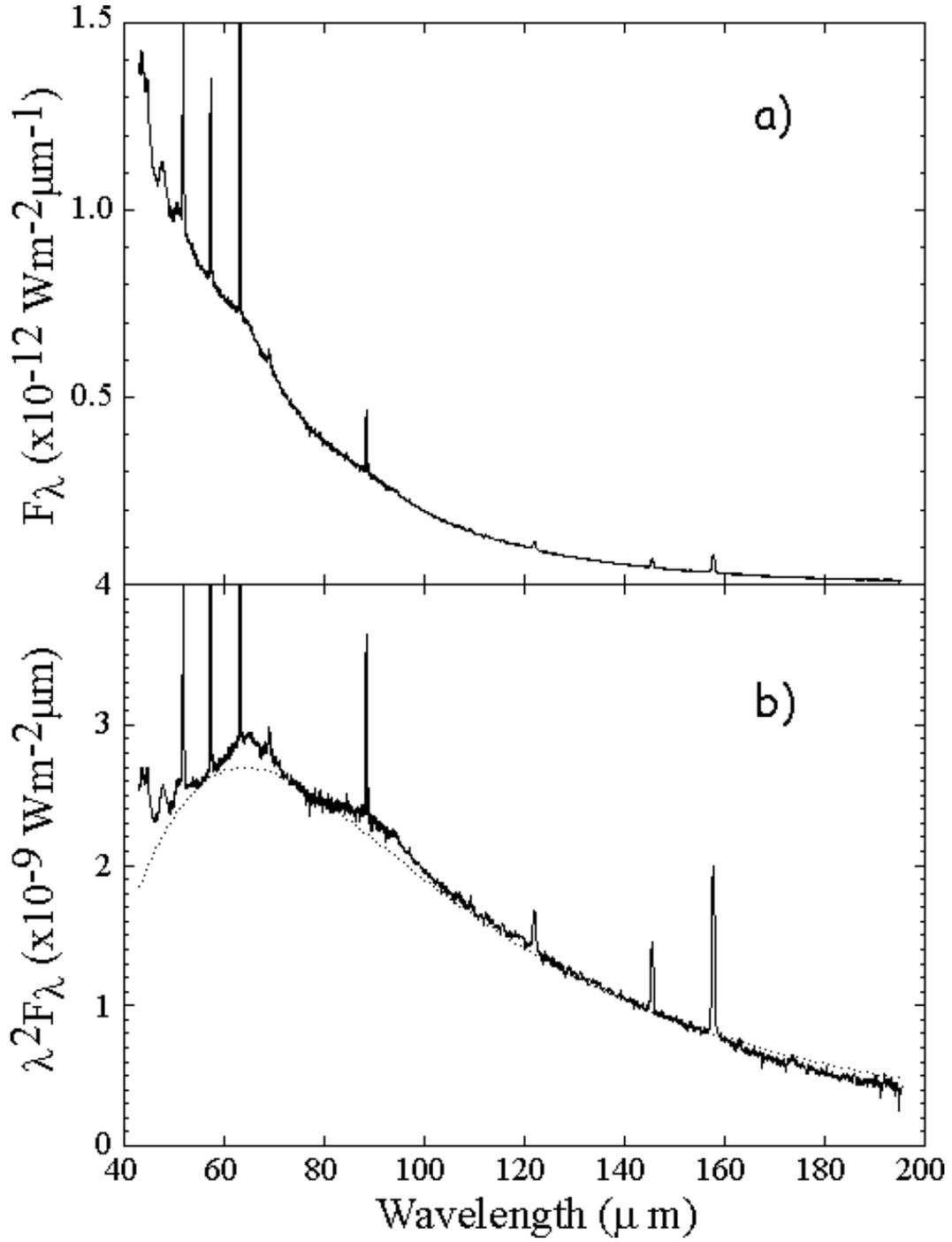}
\caption{Spectrum of NGC6302.  The dotted line indicates a graybody
of 51.8K with $\beta=1.4$ (see text).
\label{fig3}}
\end{figure}

\begin{figure}
\epsscale{1.0}

\plotone{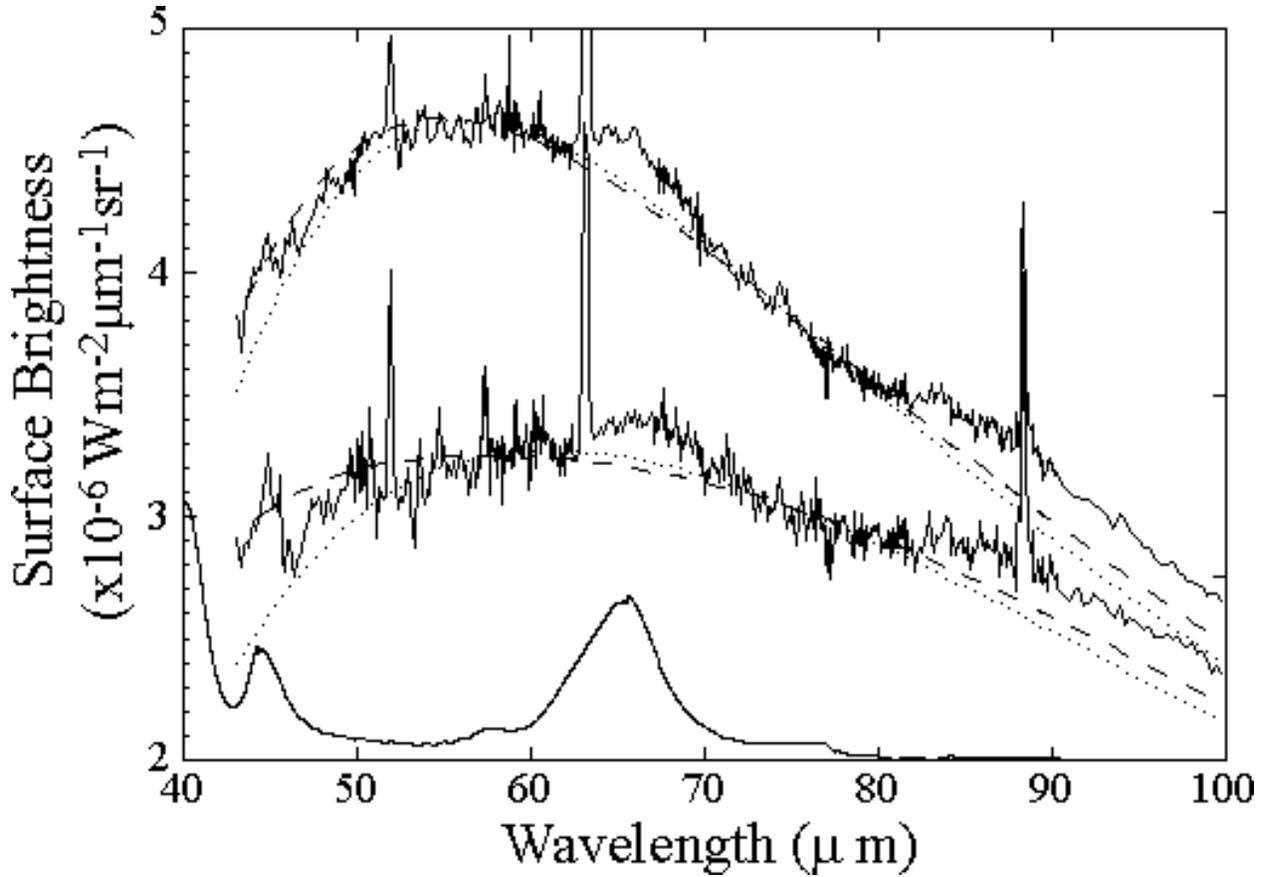}
\caption{The spectra of 40--100\,$\mu$m of the Carina (upper curve) and
S171 (middle curve) and the absorption coefficient of 
diopside at 4\,K (lower curve; Chihara et al. 2001).  The spectrum
of S171 is multiplied by 2.5.  The same best fit continua are also plotted
as in Figure~\ref{fig2}.
 \label{fig4}}
\end{figure}

\citet{kemp2002} reported the presence of
65\,$\mu$m and 90\,$\mu$m 
features in the spectra of evolved stars.  Figure~\ref{fig3}a shows
a spectrum of NGC6302 taken from the ISO Archival Data Center 
for comparison \citep[cf.][]{mol2001}.  The continuum emission indicates
a much higher temperature than those in Figure~\ref{fig2} and the 
features are weakly seen on the steep continuum.  To see the
features more clearly, the flux is multiplied by the square of
the wavelength ($\lambda^2 F_\lambda$) and plotted in Figure~\ref
{fig3}b.
The interstellar 65\,$\mu$m feature seems very similar to 
that detected in evolved stars.  The peak of the 
65\,$\mu$m feature is located obviously longer
than [\ion{O}{1}]63\,$\mu$m line (Fig.~\ref{fig4}; see also Fig.~\ref{fig6}) and 
thus is 
not compatible with the crystalline
ice band at 62\,$\mu$m \citep{smith94, mol99, coh99}.  \citet{koi2000} have 
proposed a
Ca-bearing crystalline silicate, diopside (CaMgSi$_2$O$_6$), as a possible
carrier of the 65\,$\mu$m feature in evolved stars.  Cryogenic measurements of 
the optical properties of diopside
support the identification \citep{chi2001}.  Figure~\ref{fig4} shows
a comparison of the observed spectra with the laboratory data.
The best fit continua described in next subsection are also plotted.
The laboratory spectrum shows a narrower profile than those observed
and other species, such as water ice and dolomite (CaMg(CO$_3$)$_2$),
have been suggested to contribute also to the 65\,$\mu$m band emission 
\citep{kemp2002}.  Diopside has a weak feature also
at 44.5\,$\mu$m.  The LWS detector in this spectral range (SW1) is less 
sensitive and known to have 
strong hysteresis.  In the present spectra a band feature is seen around 
45\,$\mu$m both in the upward and downward scans of both spectra, suggesting 
the presence of the 45\,$\mu$m feature.  However large noises in this spectral
range preclude the firm detection and further observations are needed to
confirm the feature.
Band features of other crystalline minerals, such as the 69\,$\mu$m
band of forsterite seen in NGC6302, are not
seen in the present LWS spectra.  Diopside also has strong features in 
30--40\,$\mu$m.
The Carina nebula was observed by Short-Wavelength
Spectrometer \citep[SWS;][]{deG96} and the spectra of 2.3--45\,$\mu$m have been
obtained.  However the SWS spectra were not taken at the same positions as
the LWS spectrum and thus the direct examination is difficult.
The SWS spectra are dominated by strong continuum and do not clearly show any 
solid bands except for the broad 22\,$\mu$m feature
\citep{chan2000}.  

\placefigure{fig3}

\placefigure{fig4}

If the identification of the interstellar 65\,$\mu$m with
diopside is correct, this is 
the first detection of a crystalline silicate in the diffuse interstellar 
medium.  
Efficient destruction of dust grains by
interstellar shocks suggests that a large fraction of interstellar dust
must be formed in interstellar space in addition to those supplied 
from stars \citep{jones96}.  Diopside is a high-temperature condensate and
may survive harsh conditions.  Calcium is a less abundant element than
magnesium or silicon, but it is highly depleted in the gas phase of the
interstellar medium \citep{sav96}.  Therefore the presence of Ca-containing dust 
should not be surprising in interstellar space.  Based on the measured band
strength of diopside \citep{chi2001} and the observed strength relative
to the continuum, we roughly estimate that 5--10\% of solar abundance calcium in 
diopside
grains is sufficient to account for the observed band emission if we
take the commonly used mass absorption coefficient of 50 cm$^2$g$^{-1}$
for the continuum emission at 100\,$\mu$m \citep{hil83}.
\citet{dem99} have suggested that the 
crystallinity of the silicates is less than 1--2\% in the interstellar medium
based on the observations of protostars.

\subsection{Interstellar 100\,$\mu$m feature}
Since the feature seen around 100\,$\mu$m is quite broad and weak, we
investigate several cases for the underlying continuum to 
examine the presence of the broad 100\,$\mu$m feature 
in detail and to make a rough estimate of the 100\,$\mu$m feature profile.  
In the estimate of the continuum we assume
the baseline positions to be at 55--60\,$\mu$m, 70--80\,$\mu$m, and
140--190\,$\mu$m or 120--190\,$\mu$m (see below)
and we try to fit the observed spectra in these ranges with the model continuum
as much as possible.  Because the shortest spectral range (55--60\,$\mu$m)
has higher noises, less weight is put on this range in the fitting.
We first adopt the dust model with the power-law emissivity ($\epsilon \propto 
\lambda^\beta$, where $\beta$ is a constant) for the continuum emission.  
We found that the single-temperature graybody model cannot
fit the entire baseline positions satisfactorily.  Particularly the 
model always gives a higher flux at long wavelengths than the observed spectra.  
This discrepancy cannot be solved by increasing $\beta$ because then the model
would provide unnecessarily large fluxes at shorter wavelengths.
Introducing a second component with a low temperature improves the fit 
drastically.  The model of $\beta=2$ both for warm and cold grains gives
reasonable fits, but still has slightly larger fluxes at 
longest wavelengths ($\lambda > 160$\,$\mu$m) than the observed spectra.  
Increase of $\beta$ from 2 to unrealistically large 3
for the cold grains does not improve the fit appreciably.

The power-law emissivity model has a spectral dependence of 
$F_\lambda \propto \lambda^{-(4+\beta)}$ in the Rayleigh-Jeans regime.
The discrepancy in the fit at longest wavelengths 
comes from the fact that the observed spectra have
a gradually changing power-law index.  The brightness distribution 
within the LWS beam affects the global
shape of the spectrum.  As shown in Figure~\ref{fig1} the unstitched spectra have
relatively 
large gaps in longer wavelengths ($>120$\,$\mu$m), suggesting an uncertainty
associated with the slope in this spectral range.  It also suggests a difficulty 
in defining the assumed baseline in the longer wavelengths.  
In the following we present
two cases for the baseline; one with 140--190\,$\mu$m (case A)
and 120--190\,$\mu$m (case B)
to examine the effect of the assumed baseline and
as a more realistic model we examine the astronomical silicate and
graphite grain model \citep{dra84}.
The silicate and graphite grains both have approximately a power-law 
emissivity of
$\beta \simeq 2$ in the far-infrared and this model provides slightly
better fits than 
the power-law emissivity model of $\beta = 2$.  
We present the results of the
silicate-graphite model in the following.

We assume different single temperatures for each of the
astronomical silicate and 
graphite grains and search for the
best fit temperatures.  The observed regions may
contain various temperature components of various dust grains and thus
these fits are a simple approximation for the underlying continuum.
In Figure~\ref{fig2} the best fit results are 
plotted together with the observed spectra.  
The dotted lines indicate the results for the case A baseline, which
fit the observed spectra reasonably well even in the longest wavelengths.
They overlap mostly with the observed spectra for $\lambda > 150$\,$\mu$m
in the plot. 
The dashed lines show those for the case B baseline, which have obviously higher
fluxes at longest wavelengths ($\lambda > 160$\,$\mu$m) 
than the observed spectra.  
Both cases clearly indicate
the presence of an excess feature starting around 80\,$\mu$m.  The
slope change around 80\,$\mu$m 
is steep and cannot be accounted for by extra graybodies.
The similarity of the excess profile in two different regions supports
the presence of the feature and suggests the common origin.

\citet{kemp2002} have attributed the 90\,$\mu$m feature in evolved stars 
to calcite (CaCO$_3$), a carbonate mineral.  In Figure~\ref{fig3}b we also
plot a single-temperature graybody as a simple reference continuum.
Comparison with Figure~\ref{fig2} indicates that
the 90\,$\mu$m feature in NGC6302 is narrower than
the interstellar 100\,$\mu$m feature.  The spectrum of NGC6302 
shows a clear slope change
around 100\,$\mu$m, which indicates the longer wavelength edge of
the feature.  In contrast, the interstellar spectra do not show the
clear change in the slope and suggest that the feature is extended to longer
wavelengths than the 90\,$\mu$m feature.  
The longer wavelength edge of the interstellar feature cannot be well determined.
Although the exact peak position and width of the feature 
depend on the assumed continuum and the location of the baseline,  
the interstellar 100\,$\mu$m
feature seems to be shifted to longer wavelengths and have a wider width
than the 90\,$\mu$m feature seen in evolved stars.
While carbonate grains are a likely candidate for the
90\,$\mu$m emission around evolved stars and may partly contribute to
the interstellar 100\,$\mu$m feature,
we examine the possibility of alternative species which has
a broad feature around 100\,$\mu$m for the
band carrier in the diffuse emission.
In the following we investigate whether carbon onion grains consisting of
concentric curved graphitic sheets \citep{uga92} can account for the observed 
broad 100\,$\mu$m feature or not.  

\subsection{Far-Infrared spectrum of carbon onion particles}

Graphite is an anisotropic material and has different optical properties
in the directions parallel and perpendicular to the
c-axis (the c-axis is perpendicular to the graphitic plane).  It has an
interband transition around 80\,$\mu$m in the direction perpendicular to
the c-axis \citep{phi77}.  The emission efficiency of graphite spheres
can be calculated by the so-called \case{1}{3}--\case{2}{3} approximation 
\citep{dra93},
in which the efficiencies in the perpendicular 
and parallel to the c-axis are averaged with the weight of \case{2}{3} and
\case{1}{3}, respectively.  This approximation is valid in the small particle
limit if the sphere consists of layered graphitic sheets and the
optical properties in both directions are independent.  In the graphite
sphere, the emission efficiency in the direction parallel to the c-axis is much
larger than that in the direction perpendicular to the axis in the
far-infrared region.  Therefore the interband transition feature mentioned above
is not visible in the averaged efficiency \citep{dra84}.

\begin{figure}
\epsscale{1.0}
\plotone{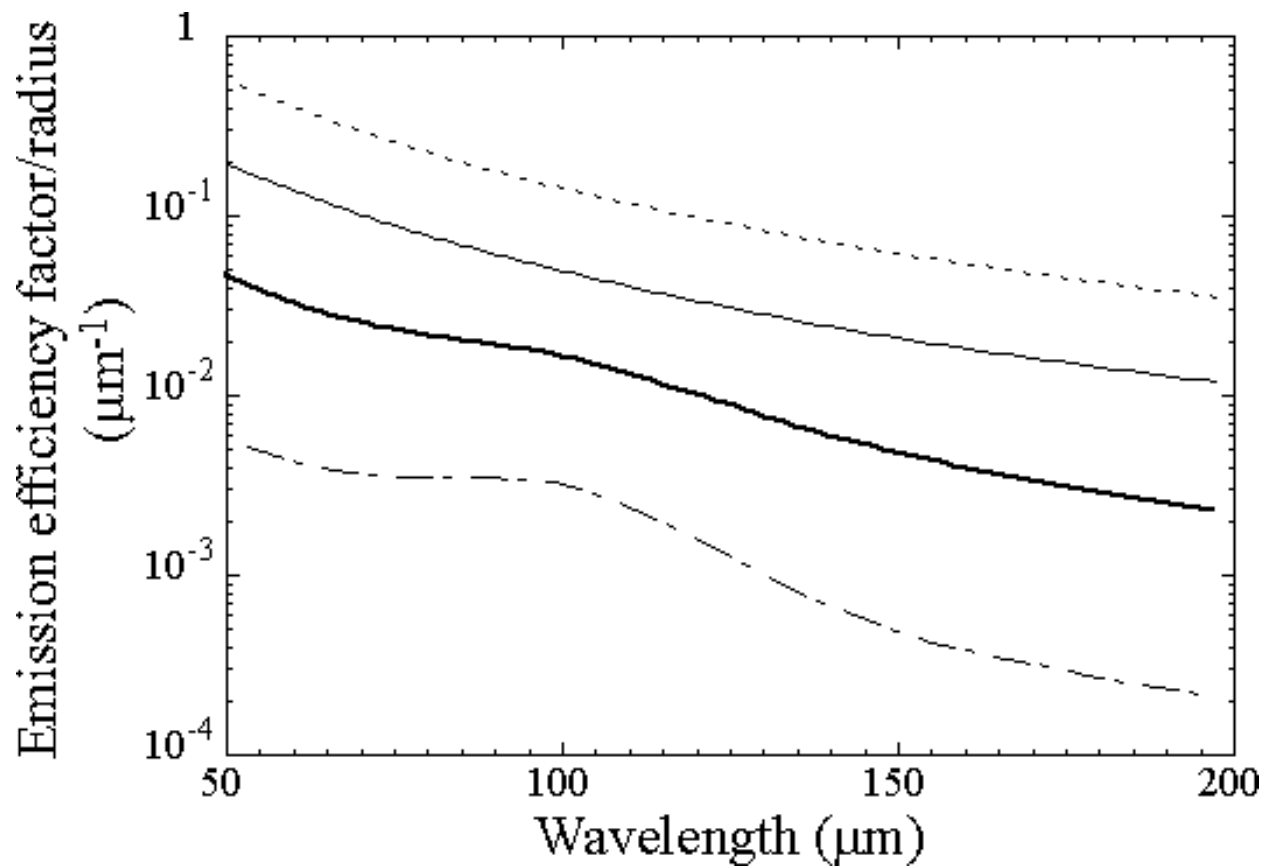}
\caption{The emission efficiency factors divided by the particle radius for 
carbon onions (thick solid line) and graphite grains (thin solid lines).  Those 
for graphite grains in the direction perpendicular to the c-axis (dot-dashed 
line) and those parallel to the c-axis (dashed line) are also plotted.  Carbon 
onion grains are assumed to have a cavity of 0.7 in radius relative to the 
particle radius.  \label{fig5}}

\end{figure}

In carbon onions, on the other hand, the graphitic layer is curved and 
approximately constitutes closed shells.  Thus the optical properties in the
both directions should be mutually coupled and the interband
feature can become visible in the emission efficiency of carbon onion
grains.  Figure~\ref{fig5} shows the emission efficiency factors divided by the
grain radius for a graphite sphere and a carbon onion grain under the assumption
that the grain radius is much smaller than the wavelengths in question.  
Here the 
dielectric constants of graphite in the directions parallel and perpendicular
to the c-axis at room temperature measurements are adopted in the calculations 
\citep[see below for discussion]{phi77, ven77}.  The
efficiency for the carbon onion is calculated by the formulation by 
Henrard et al. (1993)
and is assumed to have a central cavity of 0.7 in radius relative to
the particle size.  The appearance of the feature is insensitive to the
size of the cavity.  A broad
feature around 100\,$\mu$m is seen in the emission efficiency of the graphite
sphere in the direction perpendicular to the c-axis, but it is hardly
seen in the averaged efficiency.  On the other hand, the far-infrared
feature is evident in the emission efficiency of carbon onion particles.

\placefigure{fig5}

\begin{figure}
\epsscale{0.75}
\plotone{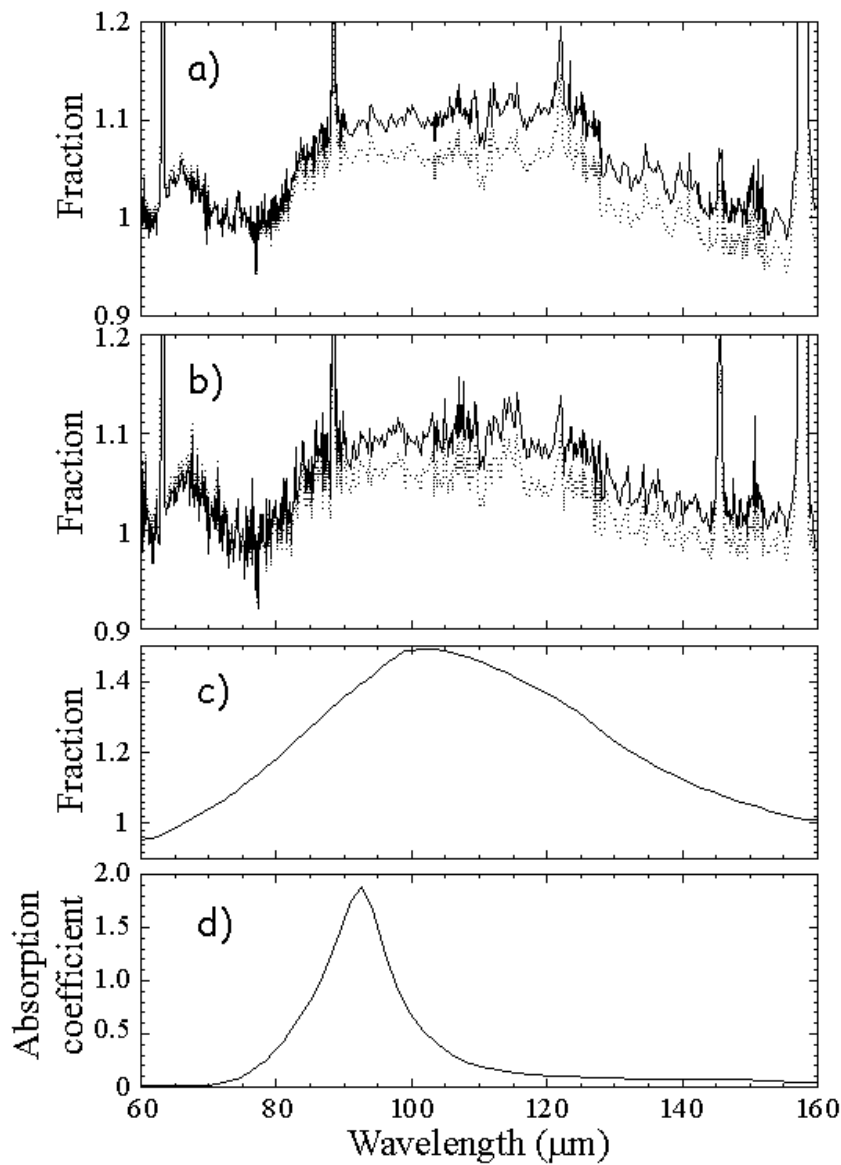}
\caption{Comparison of the carbon onion feature (c) with the observed 
interstellar feature in the Carina nebula (a) and S171 (b) together with
the absorption coefficient of calcite (d; C. Koike, 2002, private 
communication).  
The fractions of 
the spectra to the assumed continuum are plotted for (a), (b), and (c).
In (a) and (b) two lines indicate the effects of the assumed
continuum.  The solid line shows the result of case A (140--190\,$\mu$m
baseline), while the dotted line indicates that of case B (120--190\,$\mu$m
baseline) (see text).
\label{fig6}}
\end{figure}

Figure~\ref{fig6} shows a comparison of the observed feature with that of carbon 
onion grains.  To make the comparison easy the observed spectra are
divided by the assumed continuum,
while the efficiency of the carbon onion is divided by $\lambda^{-2.2}$. 
Two lines in the upper two panels indicate the effect of the assumed 
continuum.
Carbon onion grains show a similar broad feature to that observed in the
diffuse interstellar emission, but details of the profile do not match 
perfectly.  Taking account of the uncertainties in the shape of the
underlying continuum and the optical properties of carbon onions (see below),
the similarity of the band feature
suggests that carbon onions are a possible carrier of
the interstellar 100\,$\mu$m feature.

\placefigure{fig6}

The 100\,$\mu$m feature of carbon onions results from the surface resonance of
small particles \citep{bor82} and appears near the wavelength where the
real part of the dielectric constants in the perpendicular direction
just becomes below zero.  The exact position and profile of the feature thus
depend on the adopted dielectric constants.  While the electronic structure
of carbon onions has been suggested to not differ significantly from 
that of graphite \citep{pich2001}, the contribution of free electrons, which 
dominates in the
far-infrared regions, may be different.  In fact, measurements 
of electron
spin resonance and electron energy-loss spectroscopy suggest that
$\pi$ electrons in carbon onions are mostly localized in small domains
\citep{tom2001}.  The localization of $\pi$ electron will decrease the
contribution of free electrons, shifting the surface mode to wavelengths
longer than 100\,$\mu$m.  We surmise that the shift can be more than 10\,$\mu$m.
But it is difficult to estimate the possible range of the shift because
the behavior of free electrons depends also on the temperature and
the strength of the interband transition in carbon onions could also be affected 
by the localization. 
The $\pi$ electron localization, the temperature dependence,
and the possible change in the interband transition strength should affect the 
optical
properties of carbon onions in the far-infrared.  The match seen in 
Figure~\ref{fig6}
may be just a coincidence in this sense.  
The present calculation suggests that the observed band feature can
be accounted for if carbon onion grains contribute to 20--30\% of the
far-infrared emission.

Carbon is an abundant element, but the exact 
form of carbon dust
in the interstellar medium is not yet clear \cite[e.g.][]{nuth85}.
Carbon onions are a likely form other than graphite or amorphous carbon 
in addition to small aromatic particles or large molecules whose presence
has been confirmed by the infrared emission bands in the diffuse interstellar 
radiation \citep{ona96, mat96}.
Carbon onions have recently attracted attentions as a new form of
carbon material following the discovery of fullerens and their family.
In astronomy, they are suggested to be formed in interstellar 
processes \citep{uga95} and the harsh conditions accompanying interstellar
dust formation are favorable for the formation of onions \citep{hen97}. They 
have been proposed as a likely candidate for the 
interstellar
220\,nm extinction hump \citep{hen93}.  
The quenched carbonaceous composite (QCC), which
shows a feature similar to the interstellar 220\,nm hump \citep{sak83}, 
has also been
shown to contain graphitic shell structures \citep{wada99}.  
It is not unexpected that the band features of carbon onions, if exist,
also appear in the infrared region.  
In the present paper we simply
propose the possibility that the interband feature of graphite in the
far-infrared could appear in the emissivity spectrum of 
particles consisting of curved graphitic sheets and
the observed broad interstellar feature around 100\,$\mu$m may 
be accounted for by carbon onion grains.  Experimental work is definitely
needed for further investigations.

\section{Summary}

In the present paper we reported the detection of two far-infrared features
at 65\,$\mu$m and 100\,$\mu$m in the diffuse infrared emission.  The 65\,$\mu$m
band can reasonably be attributed to the Ca-rich silicate, diopside.  If this
identification is correct, this is the first detection of a crystalline
silicate feature in the interstellar diffuse emission.  The interstellar
100\,$\mu$m feature seems to be broader and peaked at longer wavelengths
than the calcite feature seen in evolved stars although the precise estimate
of the band profile is difficult.  As a possible band carrier we investigate 
the possibility that the feature originates from carbon onion grains.  While
the observed feature may be accounted for by carbon onion grains
if the assumed optical properties
are adequate, the appearance of the feature is sensitive to the electronic
structure of carbon onions.  The origin of the interstellar 100\,$\mu$m 
feature must be investigated in further experimental studies.

\acknowledgments
The authors thank K. Kawara, Y. Satoh, T. Tanab\'e, H. Okuda, T. Tsuji, 
H. Shibai, and
other members of the Japanese ISO group for their continuous help and
support.  We also thank S. Tomita and S. Hayashi for stimulating discussions
on the optical properties of carbon onions and H. Chihara and C. Koike for
providing us the far-infrared data of diopside and calcite.  
This work was supported in 
part by Grant-in-Aids for Scientific Research from the Japan Society
of Promotion of Science (JSPS).






\end{document}